\documentclass[12pt]{article}
\usepackage[dvips]{graphicx}
\usepackage{amssymb}
\usepackage{amsmath}
\newcommand{\beq}{\begin{equation}}
\newcommand{\eeq}{\end{equation}}
\newcommand{\bea}{\begin{eqnarray}}
\newcommand{\eea}{\end{eqnarray}}

\def\e2sig{e^{-2r\sigma}}

\begin{document}
\setlength{\baselineskip}{18pt}

\begin{titlepage}

\vspace*{2cm}

\begin{center}
{\Large\bf A Chiral Model of Metastable SUSY Breaking \\
\vspace*{5mm}
via Chiral/Nonchiral Seiberg Duality
} 
\end{center}
\vspace{20mm}

\begin{center}
{\large
Nobuhito Maru
}
\end{center}
%
%
%
%
\begin{center}
{\it Department of Physics, and Research and Education Center for Natural
Sciences, \\
Keio University, Hiyoshi, Yokohama 223-8521, Japan}
\end{center}
%
%
\vspace*{3cm}

\centerline{\large\bf Abstract}
\vspace*{1cm}
We propose a new chiral model of metastable supersymmetry breaking 
in the conformal window via chiral/nonchiral Seiberg duality, 
which is  focused on $Spin(7)/SU(N_f-4)$ duality in this letter. 
Following the approach of Intriligator-Seiberg-Shih model,  
a small mass perturbation in the nonchiral theory leads to 
supersymmetry breaking by rank condition in the chiral dual description. 
\end{titlepage}

Dynamical supersymmetry (SUSY) breaking provides a solution to the hierarchy problem. 
Witten index argument \cite{Witten} tells us that models with dynamical SUSY breaking (DSB) 
are restricted to chiral models, which makes the DSB model building harder. 
Several years ago, it was proposed that SUSY can be dynamically broken 
even in vector-like theories such as SUSY QCD with massive matters, known as ISS model,  
if we accept a metastablity of SUSY breaking vacuum \cite{ISS}. 
Therefore, ISS model makes DSB model building much easier. 
Also from the phenomenological viewpoint, 
the ISS model opened up an avenue to avoid Landau pole problem for QCD coupling 
in scenarios of direct gauge mediation 
where the Standard Model (SM) gauge group is embedded into a subgroup of 
unbroken flavor symmetry in DSB models.  
In the case of previously known chiral DSB models so far, 
the flavor symmetry group is typically smaller than the gauge group, 
which is likely to have a large number of messengers being charged under both groups. 
Therefore, it has been so hard to avoid the problem in chiral DSB models. 
In ISS model, the flavor symmetry group is larger than the gauge group, 
which means that the number of messengers can be reduced 
and Landau pole problem is easily solved. 

Metastable SUSY breaking is realized in the vector-like models, 
but it is interesting to ask whether this is generic or not, 
namely the metastable SUSY breaking in chiral models is possible or not. 
Along this line of thought, some studies were recently performed 
in a class of s-confining theories in SUSY gauge theories \cite{Shadmi, LMS}. 
In this paper, we propose a new chiral model of metastable SUSY breaking 
via chiral/nonchiral Seiberg duality. 
The idea is very simple. 
Let us remind that ISS model is based on massive SUSY QCD 
and SUSY breaking by rank condition is seen in the free magnetic phase description. 
Applying this procedure to chiral/nonchiral duality, 
we guess that a chiral model of metastable SUSY breaking can be easily obtained.

The model we consider is an ${\cal N}=1$ supersymmetric gauge theory 
which has a chiral/nonchiral Seiberg duality proposed by Pouliot \cite{Pouliot}. 
The nonchiral theory is a $Spin(7)$ gauge theory with $N_f$ spinor representations 
$Q(N_f, 1-\frac{5}{N_f})$ where the quantities in the parenthesis are 
representations and charges under the global symmetries $SU(N_f) \times U(1)_R$, respectively. 
There is no superpotential. 
This nonchiral theory has been known to be in the Non-Abelian Coulomb phase 
in the range $7 \le N_f \le 14$, 
and to have a {\em chiral} dual description of $SU(N_f-4)$ gauge theory 
with a symmetric tensor $s(1, \frac{2}{N_f-4})$, 
anti-fundamentals $q(\bar{N}_f, \frac{5}{N_f} - \frac{1}{N_f-4})$, 
and a singlet $M(\frac{1}{2}N_f(N_f+1), 2-\frac{10}{N_f})$. 
The quantities in the parenthesis are also 
representations and charges under the global symmetries $SU(N_f) \times U(1)_R$, respectively. 
Superpotential in the dual theory is given by
\bea
W=\frac{1}{\Lambda_m^2} Mqqs + \frac{1}{\Lambda_m^{N_f-7}}{\rm det}s 
\label{superpot}
\eea
where a dimensionful parameter $\Lambda_m$ is a cutoff scale in the dual theory. 
The gauge singlet $M$ corresponds to a meson $M=Q^2$ in the original nonchiral theory. 
The other gauge singlet operator is $B=Q^4$ for $N_f \ge 4$.

Since our interest is the fact that a metastable supersymmetry breaking takes place 
near the origin of vacuum, we introduce a mass term in the nonchiral theory as was considered 
in ISS model \cite{ISS}.  
\bea
\delta W = m Q^2~(m \ll \Lambda)
\label{mass}
\eea
where the mass $m$ is smaller than a cutoff scale of the nonchiral theory $\Lambda$. 
This mass term corresponds to a linear term of the gauge singlet in the dual theory 
\bea
\delta W = m M = m \Lambda_m \tilde{M}  
\eea
where $\tilde{M}$ is a singlet with a canonical mass dimension.

F-term conditions are derived from (\ref{superpot}) and (\ref{mass})
\bea
&&0 = \frac{\partial \tilde{W}}{\partial \tilde{M}} = \frac{1}{\Lambda_m} qqs + m \Lambda_m, 
\label{F1} \\
&&0 = \frac{\partial \tilde{W}}{\partial q} = \frac{2}{\Lambda_m} \tilde{M}qs, 
\label{F2} \\
&&0 = \frac{\partial \tilde{W}}{\partial s} = 
\frac{1}{\Lambda_m} \tilde{M}qq + \frac{1}{\Lambda_m^{N_f-7}} \frac{1}{s} {\rm det}s 
\label{F3}
\eea
where $\tilde{W}$ is a dual superpotential with the introduced mass term $\tilde{W} \equiv W+\delta W$. 

Here we are interested in the origin of moduli space as in \cite{ISS}, 
$Q=0$ in the nonchiral theory and $M=Q^2=0$ in the corresponding chiral dual theory. 
This trivially satisfies the second F-term condition (\ref{F2}). 
The third F-term condition implies ${\det}s/s=0$, 
which tells us that at least two components of symmetric tensor $s$ should be vanished 
in the diagonal basis. 
We choose them as $s_{N_f-4, N_f-4}$ and $s_{N_f-5, N_f-5}$ after diagonalization. 
Then, the first F-term condition (\ref{F1}) is reduced to
\bea
0 &=& {\rm diag} \left(
 (qqs)_{1, 1} + m_{1, 1} \Lambda_m^2, \cdots, 
 (qqs)_{N_f-6, N_f-6} + m_{N_f-6, N_f-6} \Lambda_m^2, \right. \nonumber \\
&& \left. m_{N_f-5, N_f-5} \Lambda_m^2, \cdots, m_{N_f, N_f} \Lambda_m^2 \right), 
\label{Fterm}
\eea 
where we also take the flavor diagonal masses for quarks in the nonchiral theory for simplicity.  
This condition (\ref{Fterm}) turns out to be not  satisfied 
since the rank of $qqs$ is at most $N_f-6$ but that of $m$ is $N_f$. 
Namely, SUSY is broken by rank condition at the origin of moduli space. 
The remarkable point is that this model is a chiral model in the conformal window 
unlike s-confining models in \cite{Shadmi, LMS}. 

As in previously known models of metastable SUSY breaking, 
the VEV of $\tilde{M}$ is not determined at tree level, 
but fixed by one-loop quantum correction to the potential 
since $U(1)_R$ is explicitly broken by the mass term (\ref{mass}). 
Namely, the chiral multiplet $\tilde{M}$ is a pseudo moduli.   
We numerically checked that the VEV of $\tilde{M}$ at the minimum of the potential at one-loop 
is indeed nonvanishing in the range $m/\Lambda_m \ll 1$ 
required by the longevity of lifetime of metastable SUSY breaking vacuum, 
as will be discussed soon later. 
Several examples are listed in a table below. 
\bea
\begin{array}{|c|c|}
\hline
m/\Lambda_m & \tilde{M}/\Lambda_m \\
\hline
0.73 & 2.7 \times 10^{-11} \\
0.51 & 7.7 \times 10^{-10} \\
0.34 & 0.10 \\
0.22 & 0.40 \\
0.13 & 0.62 \\
0.06 & 0.84 \\
\hline
\end{array}
\nonumber
\eea
In the dual theory of $Spin(7)$ gauge theory, a dynamical superpotential 
\bea
W_{{\rm dyn}} \propto ({\rm det}~\tilde{M})^{1/(N_f-5)}
\eea
is generated, whose form is determind by symmetries and holomorphy. 
Taking into account the dynamical superpotential, 
the first F-term condition (\ref{F1}) is modified as
\bea
0 = \frac{1}{\Lambda_m} qqs + m \Lambda_m + \Lambda_m^{\frac{2N_f-15}{N_f-5}} 
\frac{1}{N_f-5} \frac{1}{\tilde{M}} ({\rm det}\tilde{M})^{\frac{1}{N_f-5}}. 
\label{modifiedF1}
\eea
Exploiting the second F-term condition, 
the modified condition (\ref{modifiedF1}) tells us that SUSY is restored at 
\bea
\langle \tilde{M} \rangle \sim (m  \Lambda_m^{\frac{-N_f+10}{N_f-5}})^{(N_f-5)/5}.
\eea
This means that the vacuum at origin discussed above is a local minimum of the potential. 
In order for our analysis to be valid, the vacuum expectation value 
$\langle \tilde{M} \rangle$ has to be smaller than the cutoff scale $\Lambda_m$. 
\bea
\frac{\langle \tilde{M} \rangle}{\Lambda_m} \sim \left( \frac{m}{\Lambda_m} \right)^{(N_f-5)/5} \ll 1 
\Leftrightarrow \frac{m}{\Lambda_m} \ll 1. 
\eea
For this vacuum to be phenomenologically viable, its lifetime has to be sufficiently long 
compared to the age of universe. 
The lifetime roughly estimated from the bounce action \cite{DJ} has to be large, 
\bea
S_{{\rm bounce}} \simeq \frac{|\Delta \Phi|^4}{V} 
\sim \frac{|\langle \tilde{M} \rangle|^4}{|F_{\tilde{M}}|^2}
\sim \left(\frac{m}{\Lambda_m} \right)^{\frac{2}{5}(2N_f-15)} 
\eea 
where $\Delta \Phi$ is a field distance between a metastable SUSY breaking vacuum and a SUSY one. 
$V$ is a potential height of the metastable SUSY breaking vacuum. 
In estimating the potential, a canonical K\"ahler potential for $\tilde{M}$ is understood.\footnote{
As will be seen soon, the higher order K\"ahler potentials 
$(\tilde{M}^\dag \tilde{M})^n/\Lambda_m^{2n-2}~(n \ge 2)$ 
are suppressed if the longevity of the metastable vacuum is sufficiently guaranteed 
because of $\langle \tilde{M} \rangle \ll \Lambda_m$.} 
For $N_f=7$, our meta-stable DSB vacuum can be parametrically long-lived 
as long as $m/\Lambda_m \ll 1$. 
For the remaining region of the number of flavors $8 \le N_f \le 14$, 
$m/\Lambda_m \gg 1$ is required for the longevity of our meta-stable vacuum.
However, this is physically meaningless 
since the quark mass $m$ should be very larger than the cutoff scale $\Lambda_m$. 

In summary, we have proposed a new chiral model of metastable SUSY breaking 
in the conformal window via chiral/nonchiral Seiberg duality. 
A small mass perturbation in the nonchiral theory leads to 
SUSY breaking by rank condition in the chiral dual description as in ISS model. 
In this letter, we have taken a model with $Spin(7)/SU(N_f-4)$ chiral/nonchiral duality.  
Requiring the lifetime of the metastable vacuum for sufficiently long comparing to the age of universe, 
the number of flavor is constrained to be a unique value $N_f=7$. 
Although our analysis was focused on the specific model, 
our approach used in this paper can be applied to other models 
with chiral/nonchiral duality as in \cite{chiral/nonchiral} as well.

A comment on phenomenological application to direct gauge mediation is given. 
A chiral model of metastable SUSY breaking proposed in this paper has 
the flavor symmetry group $SU(N_f)$ larger than the gauge group $SU(N_f-4)$ 
unlike previously known chiral models \cite{review}. 
This feature might open up an interesting possibility to circumvent Landau pole problem for QCD coupling 
in direct gauge mediation despite the use of chiral models. 
This issue will be studied elsewhere. 

We hope that this work will shed light on the model building of SUSY breaking and its mediation. 

 \subsection*{Acknowledgments}
The author is supported in part by the Grant-in-Aid for Scientific Research  
  from the Ministry of Education, Science and Culture, Japan (21244036)
  and by Keio Gijuku Academic Development Funds.


\end{document}